\documentclass[11pt]{article}

\usepackage{fullpage}
\usepackage[utf8]{inputenc}
\usepackage{amssymb}
\usepackage[dvips]{graphicx} 
\usepackage{epsfig}
\usepackage{pstricks}

\newcommand{\Sort}{\mathrm{Sort}}
\newcommand{\Scan}{\mathrm{Scan}}
\newcommand{\Ni}{N_{\mathrm{ins}}}
\newcommand{\Nd}{N_{\mathrm{del}}}

\newtheorem{theorem}{Theorem}

\title{External Memory Three-Sided Range Reporting and Top-$k$ Queries with Sublogarithmic Updates%
\thanks{Work supported by the Danish National Research Foundation
  grant DNRF84 through the Center for Massive Data Algorithmics (MADALGO).}}

\author{Gerth Stølting Brodal \\[2ex]
  MADALGO, Department of Computer Science, Aarhus University, Denmark \\
  \texttt{gerth@cs.au.dk}}

\begin{document}

\maketitle

\begin{abstract}
  An external memory data structure is presented for maintaining a
  dynamic set of $N$ two-dimensional points under the insertion and
  deletion of points, and supporting 3-sided range reporting queries
  and top-$k$ queries, where top-$k$ queries report the $k$~points
  with highest $y$-value within a given $x$-range.  For any constant
  $0<\varepsilon\leq \frac{1}{2}$, a data structure is constructed
  that supports updates in amortized $O(\frac{1}{\varepsilon
    B^{1-\varepsilon}}\log_B N)$ IOs and queries in amortized
  $O(\frac{1}{\varepsilon}\log_B N+K/B)$ IOs, where $B$ is the
  external memory block size, and $K$ is the size of the output to the
  query (for top-$k$ queries $K$ is the minimum of $k$ and the number
  of points in the query interval). The data structure uses linear
  space. The update bound is a significant factor $B^{1-\varepsilon}$
  improvement over the previous best update bounds for the two query
  problems, while staying within the same query and space bounds.
\end{abstract}

\vspace{1ex}
\noindent\textbf{1998 ACM Subject Classification}~~E.1 Data Structures


\section{Introduction}

In this paper we consider the problem of maintaining a dynamic set of
$N$ two-dimensional points from $\mathbb{R}^2$ in external memory,
where the set of points can be updated by the insertion and deletion
of points, and where two types of queries are supported: 3-sided range
reporting queiresand top-$k$ queries. More precisely, we consider how
to support the following four operations in external memory (see
Figure~\ref{fig:queries}):
\begin{figure}[t]
  \centerline{\input{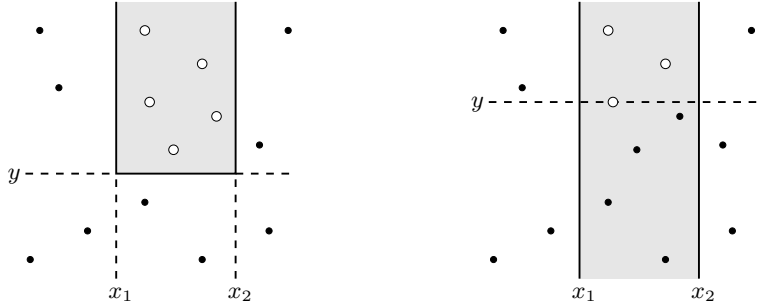}}
  \caption{3-sided range reporting queries~(left) and top-$k$
    queries~(right). The reported points are the white points and
    $k=3$.}
  \label{fig:queries}
\end{figure}
\begin{description}
\itemsep0pt
\parskip0,5ex
\item[$\mathrm{Insert}(p)$] Inserts a new point~$p\in\mathbb{R}^2$
  into the set~$S$ of points. If $p$ was already in $S$, the old copy
  of $p$ is replaced by the new copy of $p$ (this case is relevant if
  points are allowed to carry additional information).
\item[$\mathrm{Delete}(p)$] Deletes a point~$p\in\mathbb{R}^2$ from
  the current set~$S$ of points. The set remains unchanged if $p$ is
  not in the set.
\item[$\mathrm{Report}(x_1,x_2,y)$] Reports all points contained in
  $S\cap [x_1,x_2]\times[y,\infty]$.
\item[$\mathrm{Top}(x_1,x_2,k)$] Report $k$ points contained in $S\cap
  [x_1,x_2]\times[-\infty,\infty]$ with highest $y$-value.
\end{description}

\subsection{Previous work}

McCreight introduced the priority search tree~\cite{McCreight85} (for
internal memory). The classic result is that priority search trees
support updates in $O(\log N)$ time and 3-sided range reporting
queries in $O(\log N+K)$ time, where $K$ is the number of points
reported.  Priority search trees are essentially just balanced
heap-ordered binary trees where the root stores the point with minimum
$y$-value and the remaining points are distributed among the left and
right children such that all points in the left subtree have smaller
$x$-value than points in the right subtree.  Frederickson~\cite{f93}
presented an algorithm selecting the $k$ smallest elements from a
binary heap in time $O(k)$, which can be applied quite directly to a
priority search tree to support top-$k$ queries in $O(\log N+K)$ time.

Icking et~al.~\cite{wg87iko} initiated the study of adapting priority
search trees to external memory.  Their structure uses space $O(N/B)$
and supports 3-sided range reporting queries using $O(\log_2 N+K/B)$
IOs, where $B$ is the external memory block size.  Other early linear
space solutions were given in~\cite{bg90} and \cite{krvv96} supporting
queries with $O(\log_B N+K)$ and $O(\log_B N+K/B+\log_2 B)$ IOs,
respectively.  Ramaswamy and Subraminian in \cite{pods94rs} and
\cite{soda95sr} developed data structures with optimal query time and
space, respectively, but suboptimal space bounds and query bounds,
respectively (see Table~\ref{tab:results}).  The best previous dynamic
bounds are obtained by the external memory priority search tree by
Arge et al.~\cite{pods99asv}, which supports queries using $O(\log_B
N+K/B)$ IOs and updates using $O(\log_B N)$ IOs, using linear
space. The space and query bounds of~\cite{pods99asv} are
optimal. External memory top-$k$ queries were studied in
\cite{soda11abz, pods12st,pods14t}, where Tao in \cite{pods14t}
presented a data structure achieving bounds matching those of the
external memory priority search tree of Arge et al.~\cite{pods99asv},
updates being amortized.  See Table~\ref{tab:results} for an overview
of previous results.

We improve the update bounds of both \cite{pods99asv} and
\cite{pods14t} by a factor $\varepsilon B^{1-\varepsilon}$ by adopting
ideas of the buffer trees of Arge~\cite{a03} to the external memory
priority search tree~\cite{pods99asv}.

\paragraph*{1D dictionaries}

The classic B-tree of Bayer and McCreight~\cite{bm72} is the external
memory counterpart of binary search trees for storing a set of
one-dimensional points. A B-tree supports updates and
membership/predecessor searches in $O(\log_B N)$ IOs and 1D range
reporting queries in $O(\log_B N+K/B)$ IOs, where $K$ is the output
size. The query bounds for B-trees are optimal for comparison based
external memory data structures, but the update bounds are not.

Arge~\cite{a03} introduced the buffer tree as a variant of B-trees
supporting \emph{batched} sequences of interleaved updates and
queries. A sequence of $N$ operations can be performed using
$O(\frac{N}{B}\log_{M/B} \frac{N}{B})$ IOs. The buffer tree has many
applications, and can e.g.\ be used as an external memory priority
queue and segment tree, and has applications to external memory graph
problems and computational geometry problems. By adapting Arge's
technique of buffering updates (insertions and deletions) to a B-tree
of degree~$B^{\varepsilon}$, where $1>\varepsilon>0$ is a constant,
and where each node stores a buffer of $O(B)$ buffered updates, one
can achieve updates using amortized $O(\frac{1}{\varepsilon
  B^{1-\varepsilon}}\log_B N)$ IOs and member queries in
$O(\frac{1}{\varepsilon}\log_B N)$ IOs.

Brodal and Fagerberg~\cite{soda03bf} studied the trade-offs between
the IO bounds for comparison based updates and membership queries in
external memory. They proved the optimality of B-trees with buffers
when the amortized update cost is in the range $1/\log^3 N$ to
$\log_{B+1} \frac{N}{M}$.

Verbin and Zhang~\cite{vz13} and Iacono and
P\v{a}tra\c{s}cu~\cite{soda12ip} consider trade-offs between updates
and membership queries when hashing is allowed, i.e.\ elements are not
indivisible. In~\cite{soda12ip} it is proved that updates can be
supported in $O(\frac{\lambda}{B})$ IOs and queries in
$O(\log_{\lambda} N)$ IOs, for $\lambda\geq\max\{\log\log N,\log_{M/B}
(N/B)\}$. Compared to the comparison based bounds, this essentially
removes a factor $\log_B N$ from the update bounds.

\paragraph*{Related top-$k$ queries}

In the RAM model Brodal et al.~\cite{isaac09bfgl} presented a linear
space static data structure provided for the case where $x$-values
were $1,2,\ldots,N$, i.e. input is an array of $y$-values. The data
structure supports sorted top-$k$ queries in $O(k)$ time, i.e. reports
the top~$K$ in decreasing $y$-order one point at a time.

Afshani~\cite{soda11abz} studied the problem in external memory and
proved a trade-off between space and query time for sorted top~$k$
queries, and proved that data structures with query time
$\log^{O(1)}N+O(cK/B)$ requires space
$\Omega\left(\frac{N}{B}\frac{\frac{1}{c}\log_M \frac{N}{B}}{\log
  (\frac{1}{c}\log_M\frac{N}{B})}\right)$ blocks. It follows that for
linear space top-$k$ data structures it is crucial that we focus on
unsorted range queries.

Rahul et al.~\cite{walcom11rgjr} and Rahul and Tao~\cite{pods15rt}
consider the static top-$k$ problem for 2D points with associated real
weights where queries report the Top-$k$ points with respect to weight
contained in an axis-parallel rectangle. Rahul and Tao\cite{pods15rt}
achieve query time $O(\log_B N+K/B)$ using space
$O(\frac{N}{B}\frac{\log N\cdot(\log\log B)^2}{\log\log_B N})$,
$O(\frac{N}{B}\frac{\log N}{\log\log_B N})$, and $O(N/B)$ for
supporting 4-sided, 3-sided and 2-sided top-$k$ queries respectively.

\begin{table}[t]
  \newcommand{\AM}{^\dag}
  \begin{center}
    \tabcolsep5pt
    \small
    \begin{tabular}{ccccc}
      Query & Reference & Update & Query & Construction \\
      \hline
      & \cite{pods94rs} & $O(\log N\cdot\log B)\AM$ & $O(\log_B N+K/B)$ & \\
      & \cite{soda95sr} & $O(\log_B N+(\log_B N)^2/B)\AM$ & $O(\log_B N+K/B+\mathcal{IL}^*(B))$ & \\
      \raisebox{1.5ex}[0pt]{3-sided}
      & \cite{pods99asv} & $O(\log_B N)$ & $O(\log_B N+K/B)$ & \\
      & \textbf{New} & $O(\frac{1}{\varepsilon B^{1-\varepsilon}}\log_B N)\AM$ & $O(\frac{1}{\varepsilon}\log_B N+ K/B)\AM$ & $O(\Sort(N))$ \\
      \hline
      & \cite{soda11abz} &           (static)         & $O(\log_B N+ K/B)$ & \\ 
      & \cite{pods12st} & $O(\log_B^2 N)\AM$ & $O(\log_B N + K/B)$ & $O(\Sort(N))$ \\
      \raisebox{1.5ex}[0pt]{Top-$k$}          
      & \cite{pods14t} & $O(\log_B N)\AM$ & $O(\log_B N + K/B)$ &  \\
      & \textbf{New} & $O(\frac{1}{\varepsilon B^{1-\varepsilon}}\log_B N)\AM$ & $O(\frac{1}{\varepsilon}\log_B N+ K/B)\AM$ & $O(\Sort(N))$ \\
      \hline
    \end{tabular}
  \end{center}
  \caption{Previous external-memory 3-sided range reporting and
    top-$k$ data structures. All query bounds are optimal
    except~\cite{soda95sr}.  Amortized bounds are marked ``$\AM$'',
    and $\varepsilon$ is a constant satisfying $1>\varepsilon>0$. All
    data structures require space $O(N/B)$, except \cite{pods94rs}
    requiring space $O(\frac{N}{B}\log B\log\log B)$.
    $\mathcal{IL}^*(x)$ denotes the number of times $\log^*$ must be
    applied before the results becomes $\leq 2$.}
  \label{tab:results}
\end{table}

\subsection{Model of computation} 

The results of this paper are in the external memory model of Aggarwal
and Vitter~\cite{av88} consisting of a two-level memory hierarchy with
an unbounded external memory and an internal memory of size~$M$. An IO
transfers $B\leq M/2$ consecutive records between internal and
external memory. Computation can only be performed on records in
internal memory. The basic results in the model are that the scanning
and sorting an array require $\Theta(\Scan(N))$ and $\Theta(\Sort(N))$
IOs, where $\Scan(N)=\frac{N}{B}$ and $\Sort(N) =\frac{N}{B}\log_{M/B}
\frac{N}{B}$respectively~\cite{av88}.

In this paper we assume that the only operation on points is the
comparison of coordinates.  For the sake of simplicity we in the
following assume that all points have distinct $x$- and $y$-values.
If this is not the case, we can extend the $x$-ordering to the
lexicographical order $\prec_x$ where $(x_1,y_1)\prec_x(x_2,y_2)$ if
and only if $x_1<x_2$, or $x_1=x_2$ and $y_1<y_2$, and similarly for
the comparison of $y$-values.

\subsection{Our results}

This paper provides the first external memory data structure for
3-sided range reporting queries and top-$k$ queries with amortized
sublogarithmic updates.

\begin{theorem}
\label{thm:main}
  For any constant $\varepsilon$, $0<\varepsilon\leq\frac{1}{2}$,
  there exists an external memory data structure supporting the
  insertion and deletion of points in amortized
  $O(\frac{1}{\varepsilon B^{1-\varepsilon}}\log_B N)$ IOs and 3-sided
  range reporting queries and top-$k$ queries in amortized
  $O(\frac{1}{\varepsilon}\log_B N+K/B)$ IOs, where $N$ is the current
  number of points and $K$ is the size of the query output.  Given an
  $x$-sorted set of $N$ points, the structure can be constructed with
  amortized $O(N/B)$ IOs.  The space usage of the data structure is
  $O(N/B)$ blocks.
\end{theorem}

To achieve the results in Theorem~\ref{thm:main} we combine the
external memory priority search tree of Arge et al.~\cite{pods99asv}
with the idea of buffered updates from the buffer tree of
Arge~\cite{a03}.  Buffered insertions and deletions move downwards in
the priority search tree in batches whereas points with large
$y$-values move upwards in the tree in batches.  We reuse the dynamic
substructure of \cite{pods99asv} for storing $O(B^2)$ points at each
node of the priority search tree, except that we reduce its capacity
to $B^{1+\varepsilon}$ to achieve amortized $o(1)$ IOs per update.
The major technical novelty in this paper lays in the top-$k$ query
(Section~\ref{sec:top-k}) that makes essential use of Frederickson's
binary heap selection algorithm~\cite{f93} to select an approximate
$y$-value, that allows us to reduce top-$k$ queries to 3-sided range
reporting queries combined with standard selection~\cite{bfprt73}.

One might wonder if the bounds of Theorem~\ref{thm:main} are the best
possible.  Both 3-sided range reporting queries and top-$k$ queries
can be used to implement a dynamic 1D dictionary with membership
queries by storing a value $x\in\mathbb{R}$ as the 2D point
$(x,x)\in\mathbb{R}^2$. A dictionary membership query for~$x$ can then
be answered by the 3-sided query $[x_1,x_2]\times[-\infty,\infty]$ or
a top-1 query for $[x,x]$. If our queries had been worst-case instead
of amortized, it would follow from~\cite{soda03bf} that our data
structure achieves an optimal trade-off between the worst-case query
time and amortized update time for the range where the update cost is
between $1/\log^3 N$ to $\log_{B+1} \frac{N}{M}$. Unfortunately, our
query bounds are amortized and the argument does not apply.

Our query bounds are inherently amortized and it remains an open
problem if the bounds in Theorem~\ref{thm:main} can be obtained in the
worst case.  Throughout the paper we assume the amortized analysis
framework of Tarjan~\cite{tarjan85} is applied in the analysis.

\paragraph*{Outline of paper}

In Section~\ref{sec:child-structure} we describe our data structure
for point sets of size $O(B^{1+\varepsilon})$. In
Section~\ref{sec:structure} we define our general data structure. In
Section~\ref{sec:updates} we describe to how support updates, in
Section~\ref{sec:global-rebuilding} the application of global
rebuilding, and in Sections~\ref{sec:3-sided} and
Section~\ref{sec:top-k} how to support 3-sided range reporting and
top-$k$ queries, respectively. In Section~\ref{sec:construction} we
describe how to construct the data structure for a given point set.

\section{$O(B^{1+\varepsilon})$ structure}
\label{sec:child-structure}

In this section we describe a data structure for storing a set of
$O(B^{1+\varepsilon})$ points, for a constant $0\leq\varepsilon\leq
\frac{1}{2}$, that supports 3-sided range reporting queries using
$O(1+K/B)$ IOs and the batched insertion and deletion of $s\leq
B$~points using amortized $O(1+s/B^{1-\varepsilon})$ IOs. The
structure is very much identical to the external memory priority
search structure of Arge et al.~\cite[Section~3.1]{pods99asv} for
handling $O(B^2)$ points.  The essential difference is that we reduce
the capacity of the data structure to obtain amortized $o(1)$ IOs per
update, and that we augment the data structure with a sampling
operation required by our top-$k$ queries. A sampling intuitively
selects the $y$-value of approximately every $B$th point with respect
to $y$-value within a query range $[x_1,x_2]\times[-\infty,\infty]$
and takes $O(1)$ IOs.

In the following we describe how to support the below operations
within the bounds stated in Theorem~\ref{thm:child-structure}.
\begin{description}
\itemsep0pt
\parskip0,5ex
\item[$\mathrm{Insert}(p_1,\ldots,p_s)$] Inserts the points
  $p_1,\ldots,p_s$ into the structure, where $1\leq s\leq B$.
\item[$\mathrm{Deletes}(p_1,\ldots,p_s)$] Deletes the points
  $p_1,\ldots,p_s$ from the structure, where $1\leq s\leq B$ .
\item[$\mathrm{Report}(x_1,x_2,y)$] Reports all points within the
  query range $[x_1,x_2]\times[y,\infty]$.
\item[$\mathrm{Sample}(x_1,x_2)$] Returns a decreasing sequence of
  $O(B^{\varepsilon})$ $y$-values $y_1\geq y_2\geq \cdots$ such that
  for each $y_i$ there are between $iB$ and $iB+\alpha B$ points in
  the range~$[x_1,x_2]\times[y_i,\infty]$, for some constant
  $\alpha\geq 1$. Note that this implies that in the range
  $[x_1,x_2]\times[y_{i+1},y_i[$ there are between 0 and $(1+\alpha)B$
      points.
\end{description}

\begin{theorem}
\label{thm:child-structure}
  There exists a data structure for storing $O(B^{1+\varepsilon})$
  points, $0\leq \varepsilon\leq \frac{1}{2}$, where the insertion and
  deletion of $s$ points requires amortized $O(1+s/B^{1-\varepsilon})$
  IOs.  Report queries use $O(1+K/B)$ IOs, where $K$ is the number of
  points returned, and Sample queries use $O(1)$ IOs. Given an
  $x$-sorted set of $N$ points, the structure can be constructed with
  $O(N/B)$ IOs.  The space usage is linear.
\end{theorem}

\paragraph*{Data structure}

Our data structure $\mathcal{C}$ consists of four parts. A static data
structure $\mathcal{L}$ storing $O(B^{1+\varepsilon})$ points; two
buffers $\mathcal{I}$ and $\mathcal{D}$ of delayed insertions and
deletions, respectively, each containing at most $B$ points; and a set
$\mathcal{S}$ of $O(B)$ sampled $y$-values.  A point can appear at
most once in $\mathcal{I}$ and~$\mathcal{D}$, and at most in one of
them.  Initially all points are stored in $\mathcal{L}$, and
$\mathcal{I}$ and $\mathcal{D}$ are empty.

\begin{figure}[t]
  \centerline{\input{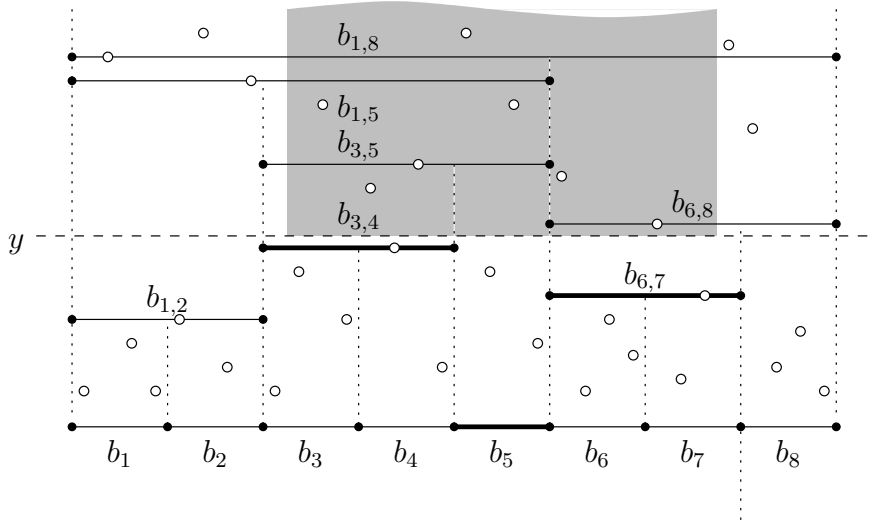}}
  \caption{$O(B^{1+\varepsilon})$ structure for $B=4$. White nodes are
    the points. Horizontal line segments with black endpoints
    illustrate the blocks stored. Each block stores the $B$ points on
    and above the line segment.}
  \label{fig:child-structure}
\end{figure}

Let $L$ be the points in the $\mathcal{L}$ structure and let
$\ell=\lceil |L|/B\rceil$.  The data structure $\mathcal{L}$ consists
of $2\ell-1$ blocks. The points in $L$ are first partitioned
left-to-right with respect to $x$-value into blocks
$b_1,\ldots,b_\ell$ each of size $B$, except possibly for the
rightmost block~$b_\ell$ just having size $\leq B$. Next we make a
vertical sweep over the points in increasing $y$-order. Whenever the
sweepline reaches a point in a block where the block together with an
adjacent block contains exactly $B$ points on or above the sweepline,
we replace the two blocks by one block only containing these $B$
points. Since each such block contains exactly the points on or above
the sweepline for a subrange $b_i,\ldots,b_j$ of the initial blocks,
we denote such a block $b_{i,j}$.  The two previous blocks are stored
in $\mathcal{L}$ but are no longer part of the vertical sweep.  Since
each fusion of adjacent blocks causes the sweepline to intersect one
block less, it follows that at most $\ell-1$ such blocks can be
created.
Figure~\ref{fig:child-structure} illustrates the constructed blocks,
where each constructed block is illustrated by a horizontal line
segment, and the points contained in the block are exactly all the
points on or above the corresponding line segment.
Finally, we have a ``catalog'' storing a reference to each of the
$2\ell-1$ blocks of $\mathcal{L}$.  For a block $b_i$ we store the
minimum and maximum $x$-values of the points within the block. For
blocks~$b_{i,j}$ we store the interval $[i,j]$ and the minimum
$y$-value of a point in the block, i.e.\ the $y$-value where the sweep
caused block~$b_{i,j}$ to be created.

The set $\mathcal{S}$ consists of the $\lceil i\cdot
B^{\varepsilon}\rceil$-th highest $y$-values in each of the blocks
$b_1,\ldots,b_\ell$ for $1\leq i\leq B^{1-\varepsilon}$. Since
$\ell=O(B^\varepsilon)$, the total number of points in $\mathcal{S}$
is $O(B^\varepsilon\cdot B^{1-\varepsilon})=O(B)$.
The sets $\mathcal{S}$, $\mathcal{I}$, $\mathcal{D}$ and the catalog
are stored in $O(1)$ blocks.

\paragraph*{Updates}

Whenever points are inserted or deleted we store the delayed updates
in $\mathcal{I}$ or $\mathcal{D}$, respectively.  Before adding a
point $p$ to $\mathcal{I}$ or $\mathcal{D}$ we remove any existing
occurrence of $p$ in $\mathcal{I}$ and~$\mathcal{D}$, since the new
update overrides all previous updates of~$p$.  Whenever $\mathcal{I}$
or $\mathcal{D}$ overflows, i.e.\ gets size $>B$, we apply the updates
to the set of points in $\mathcal{L}$, and rebuild $\mathcal{L}$ for
the updated point set.
To rebuild~$\mathcal{L}$, we extract the points $L$ in $\mathcal{L}$
in increasing $x$-order from the blocks $b_1,\ldots,b_\ell$ in
$O(\ell)$ IOs, and apply the $O(B)$ updates in $\mathcal{I}$ or
$\mathcal{D}$ during the scan of the points to achieve the updated
point set~$L'$. We split $L'$ into new blocks $b_1,\ldots,b_{\ell'}$
and perform the vertical sweep by holding in internal memory a
priority queue storing for each adjacent pair of blocks the $y$-value
where the blocks potentially should be fusioned.  This allows the
construction of each of the remaining blocks~$b_{i,j}$ of
$\mathcal{L}$ in $O(1)$ IOs per block.  The reconstruction takes
worst-case $O(\ell')$ IOs. Since $|L|=O(B^{1+\varepsilon})$ and the
reconstruction of $\mathcal{L}$ whenever a buffer overflow occurs
requires $O(|L|/B)=O(B^\varepsilon)$ IOs, the amortized cost of
reconstructing $\mathcal{L}$ is $O(1/B^{1-\varepsilon})$ IOs per
buffered update.

\paragraph*{3-sided reporting queries}

For a 3-sided range reporting query $Q=[x_1,x_2]\times[y,\infty]$, the
$t$ line segments immediately below the bottom segment of the query
range~$Q$ correspond exactly to the blocks intersected by the sweep
when it was at $y$, and the blocks contain a superset of the points
contained in $Q$. In Figure~\ref{fig:child-structure} the grey area
shows a 3-sided range reporting query $Q=[x_1,x_2]\times[y,\infty]$,
where the relevant blocks are $b_{3,4}$, $b_5$ and $b_{6,7}$. By
construction we know that at the sweepline two consecutive blocks
contain at least $B$ points on or above the sweepline. Since the
leftmost and rightmost of these blocks do not necessarily contain any
points from $Q$, it follows that the output to the range query $Q$ is
at least $K\geq B\lfloor(t-2)/2\rfloor$. The relevant blocks can be
found directly from the catalog using $O(1)$ IOs and the query is
performed by scanning these $t$ blocks, and reporting the points
contained in~$Q$.  The total number of IOs becomes $O(1+t)=O(1+K/B)$.

\paragraph*{Sampling queries}

To perform a sampling query for the range $[x_1,x_2]$ we only consider
$\mathcal{L}$, i.e.\ we ignore the $O(B)$ buffered updates.  We first
identify the two blocks $b_i$ and $b_j$ spanning $x_1$ and~$x_2$,
respectively, by finding the predecessor of $x_1$ (successor of $x_2$)
among the minimum (maximum) $x$-values stored in the catalog.
The sampled $y$-values in $\mathcal{S}$ for the blocks
$b_{i+1},\ldots,b_{j-1}$ are extracted in decreasing $y$-order, and
the $\lceil (s+1)\cdot B^{1-\varepsilon}\rceil$-th $y$-values are
returned from this list for $s=1,2,\ldots$. Let $y_1\geq y_2 \geq
\cdots$ denote these returned $y$-values.

We now bound the number of points in $\mathcal{C}$ contained in the
range $Q_s=[x_1,x_2]\times[y_s,\infty]$. By construction there are
$\lceil (s+1)\cdot B^{1-\varepsilon}\rceil$ $y$-values $\geq y_s$ in
$\mathcal{S}$ from points in $b_{i+1}\cup\cdots\cup b_{j-1}$.
In each~$b_t$ there are at most $\lceil B^\varepsilon\rceil$ points
vertically between each sampled $y$-value in~$\mathcal{S}$. Assume
there are $n_t$ sampled $y$-values $\geq y_s$ in $\mathcal{S}$ from
points in $b_t$, i.e.\ $n_{i+1}+\cdots+n_{j-1} = \lceil (s+1)\cdot
B^{1-\varepsilon}\rceil$. The number of points in $b_t$ with $y$-value
$\geq y_s$ is at least $\lceil n_t B^\varepsilon\rceil$ and less than
$\lceil (n_t+1) B^\varepsilon\rceil$,
implying that the total number of points in $Q_s\cap (b_{i+1}\cup\cdots\cup b_{j-1})$ is at least 
$\sum_{t=i+1}^{j-1} \lceil n_tB^\varepsilon\rceil\geq 
B^\varepsilon \sum_{t=i+1}^{j-1} n_t =
B^\varepsilon\lceil (s+1)\cdot B^{1-\varepsilon}\rceil\geq (s+1)B$ and at most 
$\sum_{t=i+1}^{j-1} (n_t+1)B^\varepsilon
= (j-i-1)B^\varepsilon+B^\varepsilon\sum_{t=i+1}^{j-1} n_t
= (j-i-1)B^\varepsilon+B^\varepsilon\lceil (s+1)\cdot B^{1-\varepsilon}\rceil
\leq (j-i)B^\varepsilon+(s+1)B$.
Since the buffered deletions in $\mathcal{D}$ at most cancel $B$
points from $\mathcal{L}$ it follows that there are at least
$(s+1)B-B=sB$ points in the range $Q_s$. Since there are most $B$
buffered insertions in $\mathcal{I}$ and $B$ points in each of the
blocks~$b_i$ and~$b_j$, it follows that $Q_s$ contains at most
$(j-i)B^\varepsilon+(s+1)B+3B=sB+O(B)$ points, since
$j-i=O(B^\varepsilon)$ and $\varepsilon\leq \frac{1}{2}$.  It follows
that the generated sample has the desired properties.

Since the query is answered by reading only the catalog and
$\mathcal{S}$, the query only requires $O(1)$ IOs. Note that the
returned $y$-values might be the $y$-values of deleted points by
buffered deletions in $\mathcal{D}$.

\section{The data structure}
\label{sec:structure}

To achieve our main results, Theorem~\ref{thm:main}, we combine the
external memory priority search tree of Arge et al.~\cite{pods99asv}
with the idea of buffered updates from the buffer tree of
Arge~\cite{a03}.  As in~\cite{pods99asv}, we have at each node of the
priority search tree an instance of the data structure of
Section~\ref{sec:child-structure} to handle queries on the children
efficiently. The major technical novelty lays in the top-$k$ query
(Section~\ref{sec:top-k}) that makes essential use of Frederickson's
binary heap selection algorithm~\cite{f93} and our samplings from
Section~\ref{sec:child-structure}.

\paragraph*{Structure}

The basic structure is a B-tree~\cite{bm72} $T$ over the $x$-values of
points, where the degree of each internal node is in the range
$[\Delta/2,\Delta]$, where $\Delta=\lceil B^\varepsilon\rceil$, except
for the root~$r$ that is allowed to have degree in the range
$[2,\Delta]$.
Each node $v$ of $T$ stores three buffers containing $O(B)$ points: a
\emph{point buffer} $P_v$, an \emph{insertion buffer}~$I_v$, and a
\emph{deletion buffer}~$D_v$.  The intuitive idea is that $T$ together
with the $P_v$ sets form an external memory priority search tree,
i.e.\ a point in $P_v$ has larger $y$-value than all points in $P_w$
for all descendants $w$ of $v$, and that the $I_v$ and $D_v$ sets are
delayed insertions and deletions on the way down through $T$ that we
will handle recursively in batches when buffers overflow. A point
$p\in I_v$ ($p\in D_v$) should eventually be inserted in (deleted
from) one of the $P_w$ buffers at a descendant~$w$ of $v$.
Finally for each internal node~$v$ with children $c_1,\ldots,c_\delta$
we will have a data structure $\mathcal {C}_v$ storing
$\cup_{i=1}^{\delta} P_{c_i}$, that is an instance of the data
structure from Section~\ref{sec:child-structure}.
In a separate block at $v$ we store for each child $c_i$ the minimum
$y$-value of a point in $P_{c_i}$, or $+\infty$ if $P_{c_i}$ is empty.
We assume that all information at the root is kept in internal memory,
except for~$\mathcal{C}_r$.

\paragraph*{Invariants}

For a node $v$, the buffers $P_v$, $I_v$ and $D_v$ are disjoint and
all points have $x$-values in the $x$-range spanned by the
subtree~$T_v$ rooted at~$v$ in $T$. All points in $I_v\cup D_v$ have
$y$-value less than the points in $P_v$. In particular leaves have
empty $I_v$ and $D_v$ buffers. If a point appears in a buffer at a
node~$v$ and at a descendant $w$, the update at $v$ is the most
recent.

The sets stored at a node~$v$ must satisfy one of the below size
invariants, guaranteeing that either $P_v$ contains at least $B/2$
points, or all insertion and deletion buffers in $T_v$ are empty and
all points~in $T_v$ are stored in the point buffer~$P_v$.
\begin{enumerate}
  \itemsep0,5ex
  \parskip0ex
\item
  $B/2 \leq |P_v| \leq B$, $|D_v| \leq B/4$, and $|I_v| \leq B$, or  
\item
  $|P_v|<B/2$, $I_v=D_v=\emptyset$, and $P_w=I_w=D_w=\emptyset$ for
  all descendants $w$ of $v$ in $T$.
\end{enumerate}

\section{Updates}
\label{sec:updates}

Consider the insertion or deletion of a point $p=(p_x,p_y)$. First we
remove any (outdated) occurence of $p$ from the root buffers $P_r$,
$I_r$ and $D_r$. If $p_y$ is smaller than the smallest $y$-value in
$P_r$ then $p$ is inserted into $I_r$ or $D_r$, respectively. Finally,
for an insertion where $p_y$ is larger than or equal to the smallest
$y$-value in $P_r$ then $p$ is inserted into $P_r$. If $P_r$
overflows, i.e.\ $|P_r|=B+1$, we move a point with smallest $y$-value
from $P_r$ to $I_r$.

During the update above, the $I_r$ and $D_r$ buffers might overflow,
which we handle by the five steps described below: (\textit{i}) handle
overflowing deletion buffers, (\textit{ii}) handle overflowing
insertion buffers, (\textit{iii}) split leaves with overflowing point
buffers, (\textit{iv}) recursively split nodes of degree $\Delta+1$,
and (\textit{v}) recursively fill underflowing point buffers. For
deletions only (\textit{i}) and (\textit{v}) are relevant, whereas for
insertions (\textit{ii})--(\textit{v}) are relevant.

(\textit{i}) If a deletion buffer $D_v$ overflows, i.e.\ $|D_v|>B/4$,
then by the pigeonhole principle there must exist a child $c$ where we
can push a subset $U\subseteq D_v$ of $\lceil|D_v|/\Delta\rceil$
deletions down to. We first remove all points in $U$ from $D_v$,
$I_c$, $D_c$, $P_c$, and $\mathcal{C}_v$. Any point~$p$ in $U$ with
$y$-value larger than or equal to the minimum $y$-value in $P_c$ is
removed from $U$ (since the deletion of $p$ cannot cancel further
updates). If $v$ is a leaf, we are done. Otherwise, we add the
remaining points in $U$ to $D_c$, which might overflow and cause a
recursive push of buffered deletions.  In the worst-case, deletion
buffers overflow all the way along a path from the root to a single
leaf, each time causing at most $\lceil B/\Delta\rceil$ points to be
pushed one level down. Updating a $\mathcal{C}_v$ buffer with
$O(B/\Delta)$ updates takes amortized
$O(1+(B/\Delta)/B^{1-\varepsilon})=O(1)$ IOs.

(\textit{ii}) If an insertion buffer $I_v$ overflows, i.e.\ $|I_v|>B$,
then by the pigeonhole principle there must exist a child $c$ where we
can push a subset $U\subseteq I_v$ of $\lceil|I_v|/\Delta\rceil$
insertions down to. We first remove all points in $U$ from $I_v$,
$I_c$, $D_c$, $P_c$, and $\mathcal{C}_v$. Any point in $U$ with
$y$-value larger than or equal to the minimum $y$-value in $P_c$ is
inserted into $P_c$ and $\mathcal{C}_v$ and removed from $U$ (since
the insertion cannot cancel further updates). If $P_c$ overflows,
i.e.\ $|P_c|>B$, we repeatedly move the points with smallest $y$-value
from $P_c$ to $U$ until $|P_c|=B$.  If $c$ is a leaf all points in $U$
are inserted into $P_c$ (which might overflow), and $U$ is now
empty. Otherwise, we add the remaining points in $U$ to $I_c$, which
might overflow and cause a recursive push of buffered insertions.  As
for deletions, in the worst-case insertion buffers overflow all the
way along a path from the root to a single leaf, each time causing
$O(B/\Delta)$ points to be pushed one level down. Updating a
$\mathcal{C}_v$ buffer with $O(B/\Delta)$ updates takes amortized
$O(1+(B/\Delta)/B^{1-\varepsilon})=O(1)$ IOs.

(\textit{iii}) If the point buffer~$P_v$ at a leaf~$v$ overflows,
i.e. $|P_v|>B$, we split the leaf $v$ into two nodes $v'$ and $v''$,
and distribute evenly the points $P_v$ among $P_{v'}$ and $P_{v''}$
using $O(1)$ IOs. Note that the insertion and deletion buffers of all
the involved nodes are empty. The splitting might cause the parent to
get degree~$\Delta+1$.

(\textit{iv}) While some node~$v$ has degree $\Delta+1$, split the
node into two nodes $v'$ and $v''$ and distribute $P_v$, $I_v$ and
$D_v$ among the buffers at the nodes $v'$ and $v''$
w.r.t.\ $x$-value. Finally construct $\mathcal{C}_{v'}$
and~$\mathcal{C}_{v''}$ from the children point sets~$P_c$. In the
worst-case all nodes along a single leaf-to-root path will have to
split, where the splitting of a single node costs $O(\Delta)$ IOs, due
to reconstructing $\mathcal{C}$ structures.

(\textit{v}) While some node~$v$ has an underflowing point buffer,
i.e.\ $|P_v|<B/2$, we try to move the $B/2$ top points into $P_v$ from
$v$'s children. If all subtrees below $v$ do not store any points, we
remove all points from $D_v$, and repeatedly move the point with
maximum $y$-value from $I_v$ to $P_v$ until either $|P_v|=B$ or
$I_v=\emptyset$.  Otherwise, we scan the children's point buffers
$P_{c_1},\ldots,P_{c_\delta}$ using $O(\Delta)$ IOs to identify the
$B/2$ points with largest $y$-value, where we only read the children
with nonempty point buffers (information about empty point buffers at
the children is stored at $v$, since we store the minimum $y$-value in
each of the children's point buffer). These points $X$ are then
deleted from the children's $P_{c_i}$ lists using $O(\Delta)$ IOs and
from $\mathcal{C}_v$ using $O(B^{\varepsilon})=O(\Delta)$ IOs. All
points in $X\cap D_v$ are removed from $X$ and $D_v$ (since they
cannot cancel further updates below~$v$). For all points $p\in X\cap
I_v$, the occurrence of $p$ in $X$ is removed and the more recent
occurrence in $I_v$ is moved to $X$.  While the highest point in $I_v$
has higher $y$-value than the lowest point in $X$, we swap these two
values to satisfy the ordering among buffer points.  Finally all
remaining points in $X$ are inserted into $P_v$ using $O(1)$ IOs and
into $\mathcal{C}_u$ using $O(B^{\varepsilon})=O(\Delta)$ IOs, where
$u$ is the parent of~$v$. The total cost for pulling these up to $B/2$
points one level up in~$T$ is $O(\Delta)$ IOs. It is crucial that we
do the pulling up of points bottom-up, such that we always fill the
lowest node in the tree, which will guarantee that children always
have non-underflowing point buffers if possible. After having pulled
points from the children, we need to check if any of the children's
point buffers underflows and should be refilled.

\paragraph*{Analysis}

The tree $T$ is rebalanced during updates by the splitting of leaves
and internal nodes. We do not try to fusion nodes to handle
deletions. Instead we apply global rebuilding whenever a linear number
of updates have been performed (see
Section~\ref{sec:global-rebuilding}). A leaf $v$ will only be split
into two leaves whenever its $P_v$ buffer overflows, i.e.\ when
$|P|>B$. It follows that the total number of leaves created during a
total of $N$ insertions can at most be $O(N/B)$, implying that at most
$O(\frac{N}{\Delta B})$ internal nodes can be created by the recursive
splitting of nodes. It follows that $T$ has height
$O(\log_\Delta\frac{N}{B})=O(\frac{1}{\varepsilon}\log_B N)$.

For every $\Theta(B/\Delta)$ update, in (\textit{i}) and (\textit{ii})
amortized $O(1)$ IOs are spend on each the $O(\log_\Delta
\frac{N}{B})$ levels of $T$, i.e.\ amortized $O(\frac{\Delta}{B}
\log_\Delta \frac{N}{B})=O(\frac{1}{\varepsilon
  B^{1-\varepsilon}}\log_B N)$ IOs per update. For a sequence of $N$
updates, in (\textit{iii}) at most $O(N/B)$ leaves are created
requiring $O(1)$ IOs each and in (\textit{iv}) at most
$O(\frac{N}{B\Delta})$ non-leaf nodes are created. The creation of
each non-leaf node costs amortized $O(\Delta)$ IOs, i.e.\ in total
$O(N/B)$ IOs, and amortized $O(1/B)$ IO per update.

The analysis of (\textit{v}) is more complicated, since the recursive
filling can trigger cascaded recursive refillings. Every refilling of
a node takes $O(\Delta)$ IOs and moves $\Theta(B)$ points one level up
in the tree's point buffers (some of these points can be eliminated
from the data structure during this move). Since each point at most
can move $O(\log_\Delta \frac{N}{B})$ levels up, the total number of
IOs for the refillings during a sequence of $N$ operations is
amortized $O(\frac{N}{B}\Delta \log_\Delta \frac{N}{B})$ IOs,
i.e.\ amortized $O(\frac{1}{\varepsilon B^{1-\varepsilon}}\log_B N)$
IOs per point.
The preceding argument ignores two cases. The first case is that
during the pull up of points some points from $P_c$ and $I_v$ swap
r\^oles due to their relative $y$-values. But this does not change the
accounting, since the number of points moved one level up does not
change due to this change of r\^ ole.  The second case is when all
children of a node all together have less than $B/2$ points, i.e.\ we
do not move as many points up as promised. In this case we will move
to~$v$ all points we find at the children of~$v$, such that these
children become empty and cannot be read again before new points have
been pushed down to these nodes. We can now do a simple amortization
argument: By double charging the IOs we previously have counted for
pushing points to a child we can ensure that each node with non-empty
point buffer always has saved an IO for being emptied. It follows that
the above calculations remain valid.

\section{Global rebuilding}
\label{sec:global-rebuilding}

We adopt the technique of global rebuilding \cite[Chapter 5]{o83} to
guarantee that $T$ is balanced.  We partition the sequence of updates
into epochs. If the data structure stores $\bar{N}$ points at the
beginning of an epoch the next epoch starts after $\bar{N}/2$ updates
have been performed. This ensures that during the epoch the current
size satisfies $\frac{1}{2}\bar{N} \leq N \leq \frac{3}{2}\bar{N}$,
and that $T$ has height $O(\frac{1}{\varepsilon}\log_B
\frac{3\bar{N}}{2})=O(\frac{1}{\varepsilon}\log_B N)$.

At the beginning of an epoch we rebuild the structure from scratch by
construction a new empty structure and reinsert all the non-deleted
points from the previous structure. We identify the points to insert
in a top-down traversal of the $T$, always flushing the insertion and
deletion buffers of a node $v$ to its children and inserting all
points of $P_v$ into the new tree.  The insertion and deletion buffers
might temporarily have size $\omega(B)$. To be able to filter out
deleted points etc., we maintain the buffers $P_v$, $I_v$, and $D_v$
in lexicographically sorted.  Since level~$i$ (leaves being level~0)
contains at most $\frac{3\bar{N}}{2B(\Delta/2)^i}$ nodes, i.e.\ stores
$O(\frac{\bar{N}}{(\Delta/2)^i})$ points to be reported and buffered
updates to be moved $i$ levels down, the total cost of flushing all
buffers is $O(\sum_{i=0}^{\infty}
(i+1)\frac{\bar{N}}{B(\Delta/2)^i})=O(\frac{\bar{N}}{B})$ IOs.

The $O(\bar{N})$ reinsertions into the new tree can be done in
$O(\frac{\bar{N}}{\varepsilon B^{1-\varepsilon}} \log_B \bar{N})$
IOs. The $\bar{N}/2$ updates during an epoch are each charged a
constant factor amortized overhead to cover the
$O(\frac{\bar{N}}{\varepsilon B^{1-\varepsilon}} \log_B \bar{N})$ IO
cost of rebuilding the structure at the end of the epoch.

\section{3-sided range reporting queries}
\label{sec:3-sided}

Our implementation of 3-sided range reporting queries
$Q=[x_1,x_2]\times[y,\infty]$ consists of three steps: Identify the
nodes to \emph{visit} for reporting points, push down buffered
insertions and deletions between visited nodes, and finally return the
points in the query range~$Q$.

We recursively identify the nodes to visit, as the
$O(\frac{1}{\varepsilon}\log_B N)$ nodes on the two root-to-leaf
search paths in $T$ for $x_1$ and $x_2$, and all nodes $v$ between
$x_1$ and $x_2$ where all points in $P_v$ are in~$Q$.  We can check if
we should visit a node $w$ without reading the node, by comparing $y$
with the minimum $y$-value in $P_w$ that is stored at the parent of
$w$.  It follows that all points to be reported by $Q$ are contained
in the $P_v$ and $I_v$ buffers of visited nodes $v$ or point buffers
at the children of visited nodes, i.e.\ in $\mathcal{C}_v$. Note that
some of the points in the $P_v$, $I_v$ and $\mathcal{C}_v$ sets might
have been deleted by buffered updates at visited ancestor nodes.

A simple worst-case solution for answering queries would be to extract
for all visited nodes~$v$ all points from $P_v$, $I_v$, $D_v$ and
$\mathcal{C}_c$ contained in $Q$.  By sorting the
$O(K+\frac{B}{\varepsilon}\log_B N)$ extracted points (bound follows
from the analysis below) and applying the buffered updates we can
answer a query in worst-case $O(\Sort(K+\frac{B}{\varepsilon}\log_B
N))$ IOs.  In the following we prove the better bound of amortized
$O(\frac{1}{\varepsilon}\log_B N+K/B)$ IOs by charging part of the
work to the updates.

Our approach is to push buffered insertions and deletions down such
that for all visited nodes~$v$, no ancestor $u$ of $v$ stores any
buffered updates in $D_u$ and $I_u$ that should go into the subtree of
$v$. We do this by a top-down traversal of the visited nodes. For a
visited node $v$ we identify all the children to visit. For a child
$c$ to visit, let $U\subseteq D_v \cup I_v$ be all buffered updates
belonging to the $x$-range of $c$.  We delete all points in $U$ from
$P_c$, $\mathcal{C}_v$, $I_c$ and $D_c$. All updates in $U$ with
$y$-value smaller than the minimum $y$-value in $P_c$ are inserted
into $D_c$ or $I_c$, respectively. All insertions in $U$ with
$y$-value larger than or equal to the minimum $y$-value in $P_c$ are
merged with $P_c$.  If $|P_c|>B$ we move the points with lowest
$y$-values to $I_c$ until $|P_c|=B$.  We update $\mathcal{C}_v$ to
reflect the changes to $P_c$.  During this push down of updates, some
update buffers at visited nodes might get size~$>B$. We temporarily
allow this, and keep update buffers in sorted $x$-order.

The reporting step consists of traversing all visited nodes~$v$ and
reporting all points in $(P_v \cup I_v)\cap Q$ together with points in
$\mathcal{C}_v$ contained in $Q$ but not canceled by deletions in
$D_v$, i.e.\ $(Q\cap\mathcal{C}_v)\setminus D_v$.
Overflowing insertion and deletion buffers are finally handled as
described in the update section,
Section~\ref{sec:updates}~(\textit{i})--(\textit{iv}), possibly
causing new nodes to be created by splits, where the amortized cost is
already accounted for in the update analysis.
The final step is to refill the $P_v$ buffers of visited nodes, which
might have underflowed due to the deletions pushed down among the
visited nodes. The refilling is done as described in
Section~\ref{sec:updates}~(\textit{v}).

\paragraph*{Analysis}

Assume $V+O(\frac{1}{\varepsilon}\log_B N)$ nodes are visited, where
$V$ nodes are not on the search paths for $x_1$ and $x_2$. Let $R$ be
the set of points in the point buffers of the $V$ visited nodes before
pushing updates down. Then we know $|R|\geq VB/2$. The number of
buffered deletions at the visited nodes is at most
$(V+O(\frac{1}{\varepsilon}\log_B N))B/4$, i.e.\ the number of points
reported $K$ is then at least $VB/2-(V+O(\frac{1}{\varepsilon}\log_B
N))B/4=VB/4-O(\frac{B}{\varepsilon}\log_B N)$.  It follows
$V=O(\frac{1}{\varepsilon}\log_B N+K/B)$.  The worst-case IO bound
becomes $O(V+\frac{1}{\varepsilon}\log_B
N+K/B)=O(\frac{1}{\varepsilon}\log_B N+K/B)$, except for the cost of
pushing the content of update buffers done at visited nodes and
handling overflowing update buffers and underflowing point buffers.

Whenever we push $\Omega(B/\Delta)$ points to a child, the cost is
covered by the analysis in Section~\ref{sec:updates}. Only when we
push $O(B/\Delta)$ updates to a visited child, with an amortized cost
of $O(1)$ IOs, we charge this IO cost to the visited child.
Overflowing update buffers and refilling $P_v$ buffers is covered by
the cost analyzed in Section~\ref{sec:updates}. It follows that the
total amortized cost of a 3-sided range reporting query in amortized
$O(\frac{1}{\varepsilon}\log_B N+ K/B)$ IOs.

\section{Top-$k$ queries}
\label{sec:top-k}

Our overall approach for answering a top-$k$ query for the range
$[x_1,x_2]$ consists of three steps: First we find an approximate
threshold $y$-value $\bar{y}$, such that we can reduce the query to a
3-sided range reporting query. Then we perform a 3-sided range
reporting query as described in Section~\ref{sec:3-sided} for the
range $[x_1,x_2]\times[\bar{y},\infty]$. Let $A$ be the output the
three sided query.  If $|A|\leq k$ then we return $A$. Otherwise, we
select and return $k$ points from $A$ with largest $y$-value using the
linear time selection algorithm of Blum et al.~\cite{bfprt73}, that in
external memory uses $O(|A|/B)$ IOs.  The correctness of this approach
follows if $|A|\geq k$ or $A$ contains all points in the query range,
and the IO bound follows if $|A|=O(K+B\log_B N)$ and we can find
$\bar{y}$ in $O(\log_B N+ K/B)$ IOs.  It should be noted that
our~$\bar{y}$ resembles the approximate $k$-threshold used by Sheng
and Tao~\cite{pods12st}, except that we allow an additional slack of
$O(\log_B N)$.

To compute $\bar{y}$ we (on demand) construct a heap-ordered binary
tree~$\mathcal{T}$ of sampled $y$-values, where each node can be
generated using $O(1)$ IOs, and apply Frederickson's binary
heap-selection to $\mathcal{T}$ to find the $O(k/B+\log_B N)$ largest
$y$-value in $O(K/B+\log_B N)$ time and $O(K/B+\log_B N)$ IOs. This is
the returned value~$\bar{y}$.
For each node $v$ we construct a path $\mathcal{P}_v$ of $O(\Delta)$
decreasing $y$ values, consisting of the samples returned by
$\mathrm{Sample}(x_1,x_2)$ for $\mathcal{C}_v$ and merged with the
minimum $y$ values of the point buffers $P_c$, for each child~$c$
within the $x$-range of the query and where $|P_c|\geq B/2$. The root
of~$\mathcal{P}_v$ is the largest $y$-value, and the remaining nodes
form a leftmost path in decreasing $y$-value order. For each child~$c$
of $v$, the node in $\mathcal{P}_v$ storing the minimum $y$-value in
$P_c$ has as right child the root of $\mathcal{P}_c$. Finally let
$v_1,v_2,\ldots,v_t$ be all the nodes on the two search paths in $T$
for $x_1$ and $x_2$. We make a left path~$\mathcal{P}$ containing $t$
nodes, each with $y$-value $+\infty$, and let the root of
$\mathcal{P}_{v_i}$ be the right child of the $i$th node on
$\mathcal{P}$. Let $\mathcal{T}$ be the resulting binary tree. The
$\bar{y}$ value we select is the $\bar{k}=\lceil 7t+12k/B\rceil$-th
among the nodes in the binary tree~$\mathcal{T}$.

\paragraph*{Analysis}

We can construct the binary tree~$\mathcal{T}$ topdown on demand (as
needed by Frederickson's algorithm), using $O(1)$ IOs per node since
each $\mathcal{P}_v$ structure can be computed using $O(1)$ IOs.

To lower bound the number of points in $T$ contained in
$Q_{\bar{y}}=[x_1,x_2]\times[\bar{y},\infty]$, we first observe that
among the $\bar{k}$ $y$-values in $\mathcal{T}$ larger than $\bar{y}$
are the $t$ occurrences of $+\infty$, and either $\geq
\frac{1}{3}(\bar{k}-t)$ samplings from $\mathcal{C}_v$ sets or $\geq
\frac{2}{3}(\bar{k}-t)$ minimum values from $P_v$ sets. Since $s$
samplings from $\mathcal{C}_v$ ensures $sB$ elements from
$\mathcal{C}_v$ have larger values than $\bar{y}$ and the
$\mathcal{C}_v$ sets are disjoint, the first case ensures that there
are $\geq \frac{1}{3}B(\bar{k}-t)$ points from $\mathcal{C}_v$ sets in
$Q_{\bar{y}}$. For the second case each minimum $y$-value of a $P_v$
set represents $\geq B/2$ points in $P_v$ contained in $Q_{\bar{y}}$,
i.e.\ in total $\geq
\frac{B}{2}\frac{2}{3}(\bar{k}-t)=\frac{1}{3}B(\bar{k}-t)$
points. Some of these elements will not be reported, since they will
be canceled by buffered deletions. These buffered deletions can only
be stored at the $t$ nodes on the two search paths and in nodes where
all $\geq B/2$ points in $P_v$ are in $Q_{\bar{y}}$. It follows at
most $\frac{B}{4}(t+\bar{k})$ buffered deletions can be applied to
points in the $P_v$ sets, i.e.\ in total at least
$\frac{B}{3}(\bar{k}-t) - \frac{B}{4}(t+\bar{k})
=\frac{B}{12}\bar{k}-\frac{7B}{12}t
=\frac{B}{12}\lceil 7t+12k/B\rceil-\frac{7B}{12}t
\geq k$ points will be reported by the 3-sided range reporting $Q_{\bar{y}}$.

To upper bound the number of points that can be reported by
$Q_{\bar{y}}$, we observe that these points are stored in $P_v$,
$\mathcal{C}_v$ and $I_v$ buffers. There are at most $\bar{k}$ nodes
where all $\geq B/2$ points in $P_v$ are reported (remaining points in
point buffers are reported using $\mathcal{C}_v$ structures), at most
from $t+\bar{k}$ nodes we need to consider points from the insertion
buffers $I_v$, and from the at most $t+\bar{k}$ child
structures~$\mathcal{C}_v$ we report at most
$\bar{k}B+(\alpha+1)(t+\bar{k})B$ points, for some constant $\alpha
\geq 1$, which follows from the interface of the Sample operation from
Section~\ref{sec:child-structure}.  In total the 3-sided query reports
at most $\bar{k}B + (t + \bar{k})B+\bar{k}B+(\alpha+1)
(t+\bar{k})B=O(B(t+\bar{k}))=O(\frac{1}{\varepsilon}B\log_B N+k)$
points.
In the above we ignored the case where we only find $<\bar{k}$ nodes
in~$\mathcal{T}$, where we just set $\bar{y}=-\infty$ and all points
within the $x$-range will be reported.
Note that the IO bounds for finding $\bar{y}$ and the final selection
are worst-case, whereas only the 3-sided range reporting query is
amortized.

\section{Construction}
\label{sec:construction}

In this section we describe how to initialize our data structure with
an initial set of $N$ points using $O(\Sort(N))$ IOs. If the points
are already sorted with respect to $x$-value the initialization
requires $O(\Scan(N))$ IOs.

If the points are not sorted with respect to $x$-value, we first sort
all points by $x$-value using $\Sort(N)$ IOs. Next we construct a
B-tree $T$ over the $x$-values of the $N$ points using $O(\Scan(N))$
IOs, such that each leaf stores $B/2$ $x$-values (except for the
rightmost leaf storing $\leq B/2$, $x$-values) and each internal node
has degree $\Delta/2$ (except for the rightmost node at each level
having degree $\leq\Delta/2$). The $P_v$ buffers of $T$ are now filled
bottom-up, such that each buffer contains $B$ points (except if the
subtrees below all have empty $P_w$ buffers). First we store the $N$
points in the $P_v$ buffers at the leaves of $T$ from left-to-right
using $O(\Scan(N))$ IOs. The remaining levels af $T$ are processed
bottom up by recursively pulling up points. The $P_v$ buffer of a node
is filled with the $B/2$ points with largest $y$-value from the
children, by scanning all children; if a child buffer underflows,
i.e.\ gets $< B/2$ points, then we recursive refill the child's buffer
with $B/2$ points by scanning all its children. This process
guarantees that all children of a node $v$ have $\geq B/2$ points
before filling $v$ with $B/2$ points, which enables us to move the
points to $v$ before we recursively have to refill the
children. Moving $B/2$ nodes from the children to a node can be done
with $O(\Delta)$ IOs.  In a second iteration we process the nodes
top-down filling the $P_v$ buffers to contain exactly $B$ points by
moving between 0 and $B/2$ points from the children's point buffers
$P_c$ (possibly causing $P_c$ to underflow and the recursive pulling
of $B/2$ points).  All insertion and deletion buffers $I_v$ and $D_v$
are initialized to be empty, and all $\mathcal{C}_v$ structures are
constructed from its children's $P_c$ point buffers.

We now argue that the recursive filling of the $P$ buffers requires
$O(\Scan(N))$ IOs. Level~$i$ of $T$ (leaves being level 0) contains at
most $\frac{N}{B\Delta^i}$ nodes, i.e.\ the total number of points
stored at level $i$ or above is $O(\sum_{j=i}^{\infty}
B\frac{N}{B\Delta^j})=O(\frac{N}{\Delta^i})$. The number of times we
need to move $B/2$ points to level~$i$ from level~$i-1$ is then
bounded by $O(\frac{N}{\Delta^i}/\frac{B}{2})=O(\frac{N}{B\Delta^i})$,
where each move requires $O(\Delta)$ IOs. The total number of IOs for
the filling of $P_v$ buffers becomes $O(\sum_{i=1}^{\infty}
\Delta\frac{N}{B\Delta^i}) = O(\frac{N}{B} \sum_{i=0}^{\infty}
\frac{1}{\Delta^i})=O(N/B)$.

\paragraph*{Amortized analysis} 

The above considers the worst-case cost to construct an initial
structure for $N$ points. In the following we argue that the amortized
costs of the remaining operations remain unchanged during the epoch
started by the construction.  We consider a sequence of operations
containing $\Ni$ insertions and $\Nd$ deletions, starting with a newly
constructed tree containing $N$ points.

We first bound the cost of creating new nodes in $T$ during the
updates. Since each leaf in the initial tree only spans the $x$-range
of at most $B/2$ points, it follows that $\Ni$ insertions can at most
cause $2\Ni/B$ leaves to be created. Since each new leaf of $T$ can be
created using $O(1)$ IOs, the total cost of creating new leaves is
$O(\Ni/B)$. Similarly, since each internal node has initial degree
$\leq \Delta/2$, at most $O(\frac{\Ni}{\Delta B})$ internal nodes
might be created, each taking $O(\Delta)$ IOs to create, i.e.\ in
total $O(\Ni/B)$ IOs (not counting the cost of refilling point
buffers).

An overflowing insertion buffer is handled by moving
$\Theta(B/\Delta)$ buffered insertions one level down in $T$ using
$O(1)$ IOs. Since each insertion has to be moved
$O(\frac{1}{\varepsilon}\log_B N)$ levels down before it is canceled
or transforms into the insertion into a point buffer $P_v$, it follows
that the total cost of handling over flowing insertion buffers is
$O(\frac{\Ni}{B/\Delta}\frac{1}{\varepsilon}\log_B N)$ IOs. Similarly
overflowing deletion buffers are handled by moving $\Theta(B/\Delta)$
deletions one level using $O(1)$ IOs. When the deletion of a point $p$
reaches a node where $p\in P_v$ the deletion terminates after having
removed $p$ from $P_v$. This leaves a ``hole'' in the $P_v$ buffer,
that needs to be moved down by pulling up points from the children.

Each deletion potentially creates a hole and each of the
$O(\frac{\Ni}{\Delta B})$ splittings of an internal node creates $B$
holes, i.e.\ in total we need to handle $O(\Nd+\frac{\Ni}{\Delta B}B)$
holes. Since we can move up $B/2$ points, or equivalently move down
$B/2$ holes, using $O(\Delta)$ IOs, and a hole can at most be moved
down $O(\frac{1}{\varepsilon}\log_B N)$ levels before it vanishes, the
total cost of handling holes is
$O((\Nd+\frac{\Ni}{\Delta})\frac{\Delta}{B}\frac{1}{\varepsilon}\log_B
N)$ IOs.

The total cost of handling the updates, also covering the work done by
the queries that we charged to the updates, becomes
$O(\frac{\Nd+\Ni}{B/\Delta}\frac{1}{\varepsilon}\log_B
N)=O(\frac{\Nd+\Ni} {\varepsilon B^{1-\varepsilon}}\log_B N)$ IOs,
i.e.\ matching the previous proved amortized bounds.

\bibliographystyle{plain}
\bibliography{top-k}

\newpage

\begin{appendix}
	
\section{Notation}

\begin{tabular}{cl}
  Symbol & Usage \\
  \hline
  $S$ & Current point set \\
  $N$ & Number of points, $N=|S|$ \\
  $\bar{N}$ & Number of points at start of epoch \\
  $k$ & Top-$k$ query \\
  $K$ & Output size, $K\leq k$ \\
  $Q$ & Query region \\
  $x_1,x_2,y$ & 3-sided query $Q=[x_1,x_2]\times[y,\infty]$ \\
  $d$ & Dimension, $\mathbb{R}^d$ \\
  \hline
  $B$ & Block size \\
  $M$ & Memory size \\
  \hline 
  $T$ & Base tree size \\
  $r$ & Root of $T$ \\
  $v$ & Node of $T$ \\
  $u$ & Node, parent/ancestor of $v$ \\
  $w$ & Node, descendent of $v$ \\
  $c_i$ & Node, child of  $v$ \\
  $T_v$ & Subtree rooted at $v$ \\
  \hline
  $P_v$ & Point buffer \\
  $I_v$ & Insertion buffer \\
  $D_v$ & Deleton buffer \\
  $\varepsilon$ & Construction parameter $0<\varepsilon<\frac{1}{2}$ \\
  $\delta$ & Degree of node, $\delta\leq\Delta$ \\
  $\Delta$ & Degree parameter of $T$, $\Delta=\lceil B^\varepsilon\rceil$ \\
  $p$ & Point $p=(p_x,p_y)$ \\  
  $X$ & Set points to be pulled up one level \\
  $U$ & Set of updates to be pushed down one level \\
  \hline
  $\mathcal{C}$ & Child structure \\
  $\mathcal{L}$ & List structure (child structure) \\
  $\mathcal{I}$ & Insertion buffer (child structure) \\
  $\mathcal{D}$ & Deletion buffer (child structure) \\
  $\mathcal{S}$ & Samples (child structure) \\
  $L$ & The points in $\mathcal{L}$ \\
  $\ell$ & $\ell=\lceil|\mathcal{L}|/B\rceil$ \\
  $y_i$ & Sample $y_1>y_2>\cdots$ \\
  $\alpha$ & Sample error \\
  \hline
  $\bar{y}$ & Approximate $y$-value for top-$k$ \\
  $\bar{k}$ & Parameter for Frederickson's algorithm 	\\
  $\mathcal{T}$ & Binary tree for selection of $\bar{y}$ \\
  $\mathcal{P}_v$ & Left path in $\mathcal{T}$, for node $v$ in $T$ \\
  \hline
  $i,j,s,t$ & indexes \\
\end{tabular}

\end{appendix}

\end{document}